\documentclass[RNAAS]{aastex63}

\begin{document}

\title{Supernova X-Ray Database (SNaX) Updated to Ensure Long-term Stability}

\correspondingauthor{Vikram V. Dwarkadas}
\email{vikram@astro.uchicago.edu}

\author{Alexandra Nisenoff}
\affiliation{Dept.~of Astronomy and Astrophysics \\
University of Chicago \\
5640 S Ellis Ave., Chicago, IL 60637}

\author[0000-0002-4661-7001]{Vikram V. Dwarkadas}
\affiliation{Dept.~of Astronomy and Astrophysics \\
University of Chicago \\
5640 S Ellis Ave., Chicago, IL 60637}

\author{Mathias C. Ross}
\affiliation{ Department of Mechanical and Aerospace Engineering, UCLA\\
46-147K Engineering IV, Los Angeles, CA 90095}

\begin{abstract}
The Supernova X-Ray Database (SNaX) was established  a few years ago to make X-ray data on supernovae (SNe) publicly available via an elegant searchable web interface. The database has recently been updated to PhP7, had security updates done, and moved to a new server, ensuring its long-term stability. We urge astronomers to continue to download the data as needed for their work. Those with X-ray data on SNe are requested to upload it to the database via the easily fillable spreadsheet, making it accessible to everyone.
\end{abstract}


\section{Introduction} 
Supernovae (SNe) are fascinating objects that are observed over the entire wavelength range, from radio \citep{weileretal02} to X-rays \citep{dg12}, with even a couple of unconfirmed sources in the Fermi $\gamma$-ray  waveband \citep{fermisn, fermisn2}. They are also expected to be a source of high energy cosmic-rays \citep{marcowithetal18}. Large-scale optical surveys routinely discover hundreds of SNe each year. The availability of optical data on supernovae is widespread, specifically due to the easy accessibility of data in repositories such as WISEREP \citep{wiserep}, as well as the Open SN Catalog \citep{opencatalog}. Compared to optical SNe, the field of X-ray SNe is a relatively young field. The number of detected X-ray SNe are comparably few, although they have been growing rapidly \citep{schlegel95, il03, dg12,vvd14,drrb16} since the advent of {\it Chandra}, {\it XMM-Newton} and {\it Swift}. Total detections now exceed 60 sources, many at multiple epochs. Supernovae have been detected in the X-rays more than 6 decades after explosion \citep{sp08, rd20}. All X-ray SNe except for one belong to the core-collapse type, with their progenitors being massive stars. The one exception \citep{bocheneketal18} is a Type Ia-CSM SN, which is a Type Ia SN that shows H lines in its spectrum. The H lines are assumed to arise due to the shock wave interacting with H-rich material surrounding the SN. 

With the increasing number of X-ray detections, a need was felt for an openly accessible repository of X-ray data on SNe, and thus the  Supernova X-ray Database (SNaX) was created \citep{rd17}. A web interface to the repository was made available at {\it kronos.uchicago.edu/snax}. 

\section{The SNaX database} The database is moderated, and contains only data from published sources. The interface is searchable, and a graphical user interface is provided to make and download plots, which are highly customizable. A template allows users to easily submit data to the repository for storage and retrieval. As far as possible, fluxes and luminosities are also given in a standard waveband (0.3-8 keV) so that they can be easily compared.

The data in the database are stored in a MySQL relational database. A web interface was initially built using a combination of PHP5 and JavaSscript, with graphing functionality provided by the open-source Flot graphing library (which is built upon the jQuery JavaScript library). 

\section{Updates to SNaX} 
\begin{figure*}[!htb]
\includegraphics[width=\textwidth]{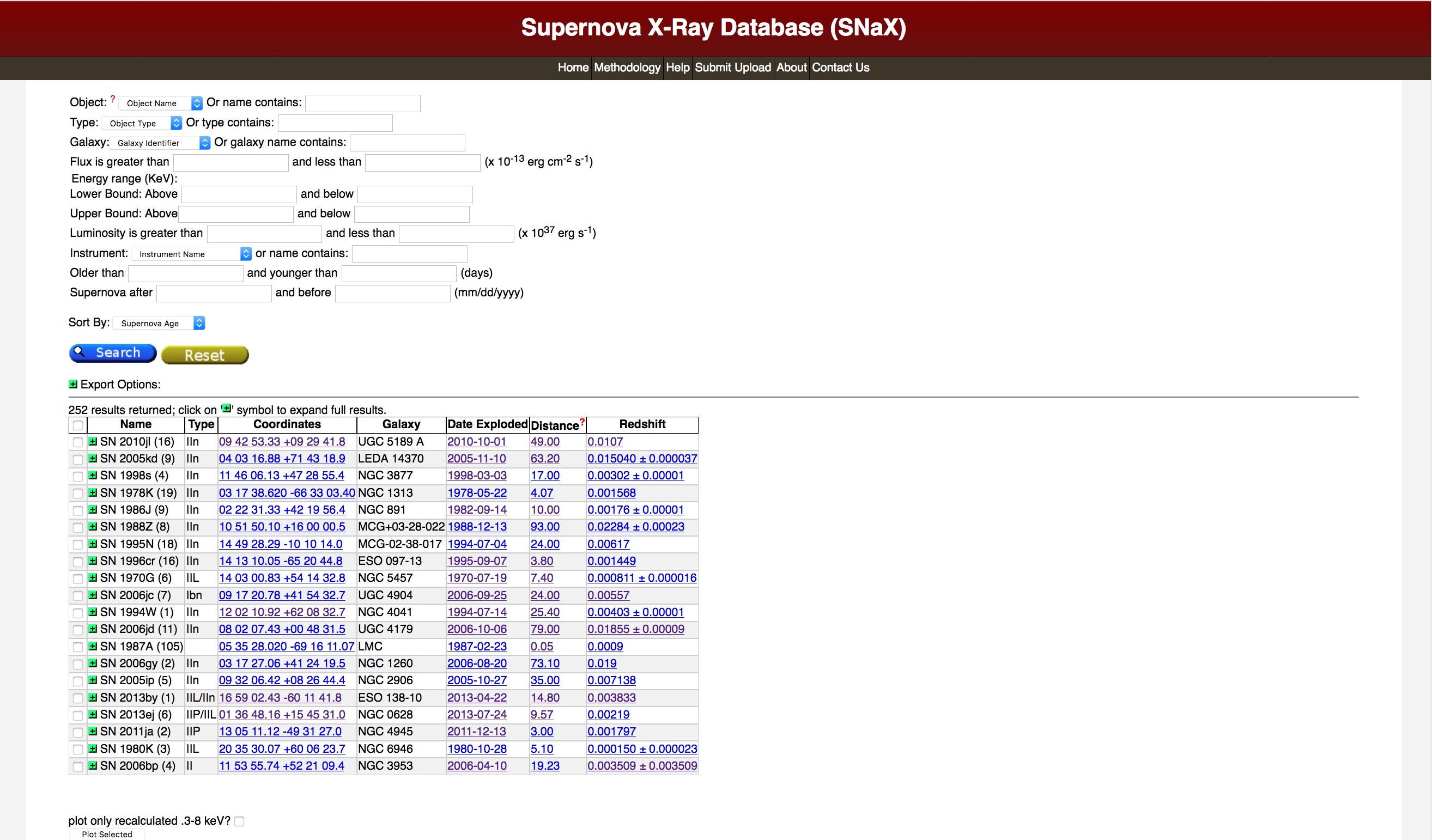}
\caption{The SNaX homepage that greets the viewer on first entering the database.
\label{fig:snax}}
\end{figure*}

Programming languages evolve over time, and machines are replaced. The SNaX database was designated to be moved to a new server, on which the version of PHP is PHP7.  This required updating of the web interface from PHP5 to PHP7. MySQL functions needed to be modified, and some scripts altered. The table on the homepage that displayed search results (Figure \ref{fig:snax}) and the administrative data verification interface on the website required updating due to incompatibilities between the two versions of PHP. Furthermore, the security of the website needed to be improved to ensure that the website could not be misused for cross-site scripting attacks. This was all duly accomplished, and we are happy to report that the updated database is now functional. SNaX is still accessible at {\it kronos.uchicago.edu/snax/} and is also now accessible at {\it supernova.uchicago.edu/snax/}. 

Users are encouraged to continue using the updated SNaX database. Please upload your X-ray data to the database, and feel free to download the data you may need in your work. In order to continue to support and improve the database, we do request users to cite the original SNaX paper \citep{rd17} as well as this Research Note, and of course the original source of the data. Please let the authors know of any problems you may encounter, and/or if you have suggestions for further improvement.

\section{Results and Conclusions}
The SNaX website has been updated to the latest version of PHP7, along with various security updates. Users of SNaX will not notice any major difference, as should  be the case if all the updates are done efficiently. Under the `hood' however a lot has changed, ensuring the long-term stability of the repository and its continued accessibility for astronomers, coupled with security updates for developers.


\acknowledgments  VVD's work is supported by NASA ADAP grant NNX14AR63G and NSF grant 1911061.

\software{SNaX} \citep{rd17},

\bibliographystyle{aasjournal} \bibliography{references}

\end{document}